\documentclass[english,11pt]{article}
\usepackage{amsmath,amsfonts,amssymb,amsbsy,amstext}
\usepackage[top=3cm, bottom=2cm, left=2cm, right=2cm]{geometry}
\usepackage{graphicx}
\usepackage[english]{babel}
\usepackage{subfigure}
\usepackage[numbers,compress]{natbib}

\newcommand{\beq}{\begin{eqnarray}}
\newcommand{\eeq}{\end{eqnarray}}
\newcommand{\non}{\nonumber\\}

\newcommand{\U}{{\rm U}}
\newcommand{\SU}{{\rm SU}}
\newcommand{\E}{{\rm E}}

\begin{document}

\begin{titlepage}

\begin{center}
{\Large\bf A Light Stop with Flavor in Natural SUSY}

\bigskip
{\large
Roberto Auzzi, Amit Giveon, Sven Bjarke Gudnason and Tomer Shacham
}

\bigskip
{\it
Racah Institute of Physics, The Hebrew University,
Jerusalem, 91904, Israel
}

\vskip 3cm

{\bf Abstract}
\end {center}

The discovery of a SM-like Higgs boson near 125 GeV and the flavor
texture of the Standard Model
motivate the investigation of supersymmetric quiver-like BSM extensions.
We study the properties of such a minimal class of models which deals
naturally with the SM parameters.
Considering experimental bounds as well as constraints from flavor
physics and Electro-Weak Precision Data, we find the following.
In a self-contained minimal model -- including the full dynamics of
the Higgs sector -- top squarks below a TeV are in tension with 
$b\to s\gamma$ constraints.
Relaxing the assumption concerning the mass generation of the
heavy Higgses, we find that a stop not far from half a TeV is allowed.
The models have some unique properties, e.g.~an enhancement of the $h\to
b\bar b,\tau\bar\tau$ decays relative to the
$h\to\gamma\gamma$ one, a gluino about 3 times heavier than the stop,
an inverted hierarchy of about $3\div 20$ between the squarks of the first
two generations and the stop, relatively light Higgsino neutralino
or stau NLSP, as well as heavy Higgses and a $W'$ which may be within
reach of the LHC.

\vfill
\noindent
\rule{5cm}{0.5pt}\\
{\it\footnotesize
 auzzi(at)phys.huji.ac.il  \\
 giveon(at)phys.huji.ac.il  \\
 gudnason(at)phys.huji.ac.il \\
 tomer.shacham(at)phys.huji.ac.il
}

\end{titlepage}

\section{Introduction}

The discovery of a SM-like Higgs boson at the LHC
\cite{ATLAS-CONF-2012-093,CMS-PAS-HIG-12-020} provides us with the
last of the eighteen SM parameters.
We take this as an opportunity to (re)consider models Beyond the
Standard Model (BSM) addressing naturally the full texture of the
SM. A perturbative Higgs near 125 GeV points towards a supersymmetric
(SUSY) extension of the SM as a possible explanation to the hierarchy
problem.
The Minimal Supersymmetric extension of the SM (MSSM), however,
requires the stop to be heavier than 5 TeV without sizable $A$-terms
(see e.g.~\cite{Espinosa:1999zm,Draper:2011aa}), in order to
radiatively generate the appropriate quartic term in the Higgs
potential.
On the other hand, such a heavy stop does not cut off the quadratic
divergences of top loops at a sufficiently low energy and,
consequently, a tuning at the per-mille level is necessary
\cite{Ellis:1986yg,Barbieri:1987fn}. This tension hints for a
supersymmetric extension with a mechanism to crank up the mass of the
lightest CP-even Higgs, $m_{h_0}$.

Furthermore, the direct search for missing transverse
energy (MET) is pushing up the bounds on the masses of the first
generation squarks to be well above the TeV
\cite{ATLAS-CONF-2012-033}. The second 
generation squarks typically need to be very close to that of the
first generation in order to pass the bounds coming from meson
mixings. The bounds on the third generation squarks, however, remain
much weaker \cite{Collaboration:2012si}. This experimental fact
together with the desire to alleviate fine tuning calls for an
inverted hierarchy of sparticle masses. This is sometimes called
effective SUSY, natural SUSY or more minimal SUSY
\cite{Dimopoulos:1995mi,Cohen:1996vb,Barbieri:2010pd,Asano:2010ut,Barbieri:2011ci,Brust:2011tb,Papucci:2011wy}.
For concrete models realizing this scenario, see
e.g.~\cite{ArkaniHamed:1997fq,Gabella:2007cp,Craig:2009hf,Aharony:2010ch,Craig:2011yk,Gherghetta:2011wc,Delgado:2011kr,Auzzi:2011eu,Csaki:2012fh,Craig:2012yd,Larsen:2012rq,Craig:2012hc,Craig:2012di,Cohen:2012rm,Randall:2012dm}.

In this paper we consider a quiver-like extension of the
Supersymmetric Standard Model (SSM), which essentially consists of two
copies of the SM gauge group ($\U(1)\times \SU(2)\times \SU(3)$) with
appropriate link fields connecting them, see
fig.~\ref{fig:quiver2nodes}. The link fields acquire a VEV via the
Higgs mechanism, breaking the gauge symmetry down to that of the SM.
If the Higgsing of the link fields takes place near a few TeV,
non-decoupling of the D-terms will contribute to the Higgs quartic
coupling at tree level. This contribution alone may allow for a 126
GeV Higgs
\cite{Batra:2003nj,Maloney:2004rc,Craig:2011yk,Auzzi:2011gh,Arvanitaki:2011ck,Auzzi:2011eu,Craig:2012hc}.

Interestingly, such extensions of the SM may also address the flavor
problem \cite{Craig:2011yk,Auzzi:2011eu,Craig:2012hc} by choosing the
messengers of SUSY breaking and the chiral superfields of the first
two generations, $q_{1,2}$, to be connected to node $B$, while the
matter of the third generation, $q_3$, and the Higgs superfields,
$H_{u,d}$, are charged under node $A$, see
fig.~\ref{fig:quiver2nodes}.
This automatically gives rise to a flavor texture in the fermion
sector, with a hierarchy between the third and the first two
generations, due to the structure of irrelevant gauge-invariant terms,
which are suppressed by the UV scale of flavor dynamics
\cite{Craig:2011yk,Auzzi:2011eu}.
The precise flavor texture depends on the representations $R$ of the
link fields $\omega,\tilde{\omega}$ in fig.~\ref{fig:quiver2nodes}.
This was analyzed in detail in \cite{Auzzi:2011eu}, where it was shown
that in several cases, the SM parameters are naturally obtained, and
the flavor constraints are satisfied.\footnote{In the simple
  construction of fig.~\ref{fig:quiver2nodes}, one needs a tuning of
  about 5 percent to generate the hierarchy between the first two
  generations, and a tuning at the level of a (few) percent of a
  couple of CP phases in the mass matrix of the soft scalars.}
Furthermore, this construction gives rise also to the above-mentioned 
inverted hierarchy between the first/second and third generation
sfermions. As the first and second generations are charged under the 
same gauge group as the messenger fields, they acquire masses
as in gauge mediation, while the third generation masses are
suppressed as in gaugino mediation
\cite{Kaplan:1999ac,Chacko:1999mi,Csaki:2001em,Cheng:2001an,McGarrie:2010kh,Green:2010ww,McGarrie:2010qr,Auzzi:2010mb,Sudano:2010vt}.

We are interested in natural models and choose to define this
statement by allowing fine tuning of UV parameters in the Lagrangian
\cite{Ellis:1986yg,Barbieri:1987fn} of only down to the percent level,
but no further. Conventionally, this concept is tightly connected to
the Higgs sector. Here we consider a broader version of the argument
where we do not allow the tuning of \emph{any} parameter in all
sectors of the Lagrangian to be tuned more than at the percent level
(using a similar definition as in the Higgs sector); this includes the
parameters describing the flavor and CP violating operators etc.
Concretely, we consider a sparticle spectrum to be natural if the
following criteria are met: the stops are lighter than about a TeV; the
gluino\footnote{Here we are working with Majorana gluini and hence we
apply the quoted naturalness bound. Dirac gluini can be twice as
heavy yielding still the same fine-tuning.} weighs less than about $3$
times the stop mass \cite{Brust:2011tb}; $\mu<200$ GeV.

The scope of this paper is to investigate -- within the class of
models described above and with the mentioned naturalness criteria --
the question of how light the stop mass can be for spectra passing all
present collider bounds, electroweak precision tests and flavor
constraints. A relatively light stop (in the ballpark of half a TeV),
can only exist if the other squarks are much heavier. Remarkably, such
an inverted hierarchy is automatic in the model of
fig.~\ref{fig:quiver2nodes}.
If the VEV of the link field $v=\langle\omega\rangle$ is smaller or of
the order of the messenger scale $M$, then the squarks of the 3rd
generation are indeed much lighter than the other squarks.
A self-contained minimal model -- including the dynamics of the Higgs
sector -- has the following properties. A light stop is tied with a
small $m_{H_d}^2$; the latter is in tension with $b\to s\gamma$
constraints. We consequently find that the stop below a TeV in the
self-contained minimal model is in tension with experiment. 
On the other hand, treating $m_{H_d}^2$ as a free parameter at the
messenger scale, we find that the stop can be accommodated in the
$600\div 1000$ GeV range.
Having the Higgs at 126 GeV as well as light stops yields
typically a ``light'' $W'$ vector boson, near $4\div 10$ TeV. 
Furthermore, the model typically has either a Higgsino neutralino NLSP
or sometimes a stau NLSP near 100 GeV.
Interestingly, this class of models has relatively heavy electroweak
gaugini making 
it possible for the self-contained version of the model to explain
\cite{Craig:2011yk} the $\mu/B\mu$ problem  providing a reasonable
$\tan\beta$ without any further dynamics in the Higgs sector.

The paper is organized as follows. In sec.~\ref{sec:overview},
we present an overview of the general properties
of the minimal construction of fig.~\ref{fig:quiver2nodes}, while
in sec.~\ref{sec:minimalmodel}, we present the details of
the model as well as the results of the paper.
In sec.~\ref{sec:unification}, we contemplate an extension that may
unify in the standard way (as opposed to accelerated
unification \cite{ArkaniHamed:2001vr} or the type studied in
\cite{Craig:2012hc}, which may be applied also to the minimal model of sec.~\ref{sec:overview}).
We conclude with a discussion, and some details are presented in the appendices.

\section{Overview of the minimal model\label{sec:overview}}

\begin{figure}[!ht]
\begin{center}
\includegraphics[width=0.35\linewidth]{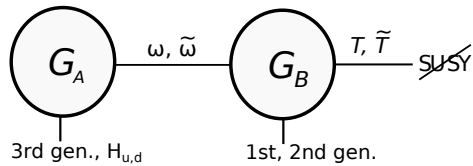}
\caption{A diagram describing the minimal model with gauge groups
$G_A,G_B=\U(1)\times\SU(2)\times\SU(3)$
  and link fields $\omega,\tilde{\omega}$. SUSY breaking is
  connected via messenger fields $T,\tilde{T}$ only to $G_B$. }
\label{fig:quiver2nodes}
\end{center}
\end{figure}
The model of BSM physics we study is characterized by the following
scales. SUSY breaking is communicated to the visible sector of the
model at the messenger scale $M$.\footnote{We will consider only
  perturbative physics throughout this paper; although so far we do
  not restrict to a specific secluded sector, eventually we consider
  messenger sectors as in minimal gauge mediation, for simplicity.} 
The visible sector consists of two copies of the SM gauge group which
are connected by certain link fields $\omega,\tilde{\omega}$, see
fig.~\ref{fig:quiver2nodes}. The link fields are chosen in
representations such that when they acquire a VEV
$\langle\omega\rangle=\langle\tilde{\omega}\rangle=v$, they Higgs the
two gauge groups down to the low-energy SM group.
Above the messenger scale $M$, we contemplate an appropriate UV
completion involving certain dynamics responsible for creating the
flavor texture of the SM, which we however only parametrize by
higher-dimension operators all suppressed by the scale
$\Lambda_{\rm flavor}\gtrsim M$. 
The models of the type we consider can have a UV completion in terms
of a deformed SQCD where $\Lambda_{\rm flavor}$ then is the strong
coupling scale of the latter theory \cite{Green:2010ww}.
We will consider the case where $\Lambda_{\rm flavor}\gtrsim M\gg v$
with $v$ near the weak scale; it is however constrained by electroweak
precision tests (EWPTs) to be typically above 1.5 TeV.

The matter content of the theory is arranged as follows. The first and
second generation matter fields are charged under the group $G_B$ (see
fig.~\ref{fig:quiver2nodes}) while the Higgses and the third
generation are charged under the group $G_A$. The superpartner masses
for the matter on the node $G_B$ are given (for $v\ll M$) as in
minimal gauge mediation 
\cite{Giudice:1998bp}, while those of $G_A$ are suppressed
by $v/M$ as in gaugino mediation
\cite{Kaplan:1999ac,Chacko:1999mi,Csaki:2001em,Cheng:2001an,McGarrie:2010kh,Green:2010ww,McGarrie:2010qr,Auzzi:2010mb,Sudano:2010vt}: 
\beq
m_{\tilde{Q}_{1,2},\tilde{u}_{1,2},\tilde{d}_{1,2},\tilde{L}_{1,2},\tilde{e}_{1,2}}
\simeq M_{1,2,3} \simeq {\rm a\ few \times}\mu_{\rm EW} \, , \qquad
m_{\tilde{Q}_{3},\tilde{u}_{3},\tilde{d}_{3},\tilde{L}_{3},\tilde{e}_{3}}
\simeq B\mu \simeq 0 \, ,
\eeq
at the messenger scale $M$. $M_{1,2,3}$ are the (heavy) gaugino masses.
The low-energy masses of the third generation and those of the Higgses
are all generated via RG evolution which in turn requires a
sufficiently heavy gluino etc. The $\mu$ term however is generated by
a higher dimension operator \cite{Craig:2011yk},\footnote{A tree-level
  $\mu_0 H_u H_d$ term is forbidden by some symmetry.}
\beq
\mathcal{W}_\mu \sim
\frac{\omega\tilde{\omega}}{\Lambda_{\rm flavor}} H_u H_d \, ,
\label{eq:muterm}
\eeq
which for $\Lambda_{\rm flavor }\approx 100$ TeV gives $\mu$ of the
right order of magnitude, viz.~of the weak scale $\mu_{\rm EW}$.

Only the third generation SM fermions receive a mass at tree level,
explaining the large top, bottom and tau mass with respect to the
other ones. The hierarchy between the top and the bottom is provided by
$\tan\beta$.\footnote{In order to explain the ratio of the bottom to
  the top mass, $\tan\beta$ needs to be of the order 50 which
  turns out to be rather high concerning experimental
  constraints. Since however we allow an up-to-a-percent-level tuning,
  this complies with our ambitions.}
The remaining part of the SM fermions acquires masses via
higher-dimension operators involving the link fields
$\omega,\tilde{\omega}$.
The representation of the link fields determines the Yukawa texture
of the SM fermions. In the simplest case the link fields are
bifundamental fields of $\SU(5)$, transforming as
$({\bf 5},\bar{\bf 5})$ under the two gauge groups $G_A\times G_B$
(which decomposes as a field
$({\bf 2},\bar{\bf 2})_{\frac12,-\frac12}$ under $\SU(2)\times\U(1)$
which we denote $\omega_L$ and as
$({\bf 3},\bar{\bf 3})_{\frac13,-\frac13}$ under
$\SU(3)\times\U(1)$ which we call $\omega_d$). The higher dimensional
representation $({\bf 10},\overline{\bf 10})$ gives rise to a somewhat
better flavor texture \cite{Auzzi:2011eu} while also adding a lot of
matter fields which can pose problems in terms of a Landau pole.
In this paper we will stick to the simplest link fields, namely those
transforming as $({\bf 5},\bar{\bf 5})$. An example of such a
higher-dimension operator is 
\beq
\frac{\lambda_{i j}^u}{\Lambda_{\rm flavor}} Q_i H_u u_j^c \omega_L \,
, \qquad i,j=1,2, \ (\rm generation\ indices) \, ,
\eeq
producing Yukawa textures schematically as
\cite{Craig:2011yk,Auzzi:2011eu}
\beq
Y_u \sim
\begin{pmatrix}
\epsilon & \epsilon & \epsilon^2 \\
\epsilon & \epsilon & \epsilon^2 \\
\epsilon^2 & \epsilon^2 & 1
\end{pmatrix} \, , \qquad
Y_d \sim
\begin{pmatrix}
\epsilon & \epsilon & \epsilon^2 \\
\epsilon & \epsilon & \epsilon^2 \\
\epsilon & \epsilon & 1
\end{pmatrix} \, , \qquad
Y_e \sim
\begin{pmatrix}
\epsilon & \epsilon & \epsilon \\
\epsilon & \epsilon & \epsilon \\
\epsilon^2 & \epsilon^2 & 1
\end{pmatrix} \, , \qquad
\epsilon \equiv \frac{v}{\Lambda_{\rm flavor}} \, ,
\label{eq:Yukawas}
\eeq
for the up, down and LH lepton sectors, respectively. The matrices
should be understood as the order of magnitude of the higher-dimension
operators, viz.~each element is multiplied by its own
order-one coefficient. An appropriate set of order-one numbers can
reproduce the measured quark masses as well as the CKM matrix
\cite{Auzzi:2011eu}.

Concerning phases we impose CP conservation at the messenger scale $M$ such that
all the gaugino masses are real valued. RG evolution does not change this.
Further phases in the Higgs sector could a priori be of concern,
however, we tackle those together with a solution to the
$\mu/B\mu$ problem.
$\mu$ is generated by the higher-dimension operator \eqref{eq:muterm}
giving roughly the 100 GeV scale (and it runs very slowly during RG),
while $B\mu$ is negligible at the scale $M$ but is generated via RG
evolution ($B\mu$ is a SUSY-breaking parameter and receives
contributions comparable to the other fields on node $G_A$, hence it is
very suppressed at the scale $M$).
$B\mu$ turns out to be rather small in other models with this type of
mechanism, however since in our model the gaugini are necessarily
rather heavy, a sufficiently large $B\mu$ (of order $\mu^2$) is
obtained, giving rise to acceptable values of $\tan\beta$.
This way of generating $B\mu$ ensures that the CPV phase 
arg$(\mu^*B\mu)$ is negligible.

As mentioned, the gluino is quite heavy in our model, due
to the fact that while the stop is very light at the messenger scale,
a sizable stop mass is needed for driving the up Higgs tachyonic
in time for EWSB to happen. This means that the gluino has to be sufficiently
heavy as to feed enough mass into the stop so EWSB occurs, but
not too heavy so fine-tuning is sufficiently small.
Another characteristic coming along with the negligible masses on the
node $G_A$ at the scale $M$ is that the down Higgs is
typically small in the model if no additional contribution is
contemplated which in turn renders the heavy Higgses rather light, of
the order of less than 400 GeV.

Finally, let us mention that even though the gluino weighs in pretty
well and we typically find the stops in the
range of $400\div 1000$ GeV, the Higgsino mass $\mu$ is generically as
low as 100 GeV  -- all in all giving rise to just a little fine tuning
in the model at hand.

Clearly, we wish to obtain the complete low-energy spectra and in
turn what predictions can be drawn from those.
Since we have
heavy link fields $\omega,\tilde{\omega}$ as well as two gauge groups
between the scales $M$ and $v$ we implement a custom made RG code that
evolves the running masses at two loops and gauge couplings at one
loop down to $v$. The light third generation receives an extra
contribution due to a threshold effect of integrating out the heavy
gaugini and link fields \cite{DeSimone:2008gm}. After Higgsing we sum
up everything and plug in the evaluated masses to the spectrum
calculator SOFTSUSY which we use to calculate the pole masses of the
particles.
Since starting with arbitrary model parameters at the messenger scale
$M$, it is not likely to provide a spectrum which is not ruled out by
experimental constraints or is not far from them. In order to put the
model on the edge of exclusion (or discovery, depending on the point
of view), we apply direct search constraints as well as electroweak
precision tests (EWPTs) to the evaluated spectrum to determine whether
it matches what we just described.
The direct search bounds we apply are from both LEP, Tevatron and the
LHC and are applied to the chargini, neutralini, gluino, first and
second generation squarks, stau, stau neutrino and CP-odd Higgs
$A_0$.
The indirect limits are applied to the charged Higgses $H_\pm$ (from
flavor measurements of $b\to s\gamma$), and the oblique $T$ parameter
limits on the VEV $v$ (up to a combination of the gauge couplings) as
well as the soft mass of the link fields. We also consider a larger
set of EWPTs setting (somewhat independent) limits also on $v$.
Now when a spectrum is calculated, we calculate an error value based
on asymmetric potentials (with large coefficients) pushing the
spectrum towards the allowed region with respect to the limits. The
bottom of this potential consists basically of the stop mass. Finally,
we use educated guesses for the starting point as well as a steepest
descent algorithm to find a spectrum with as low stop masses as
possible in a spectrum satisfying all desired constraints. This will
be presented in sec.~\ref{sec:benchmarks}.

Let us sum up what we found. As mentioned the heavy Higgses ($H_\pm$)
turn out to be generically too light with respect to the flavor
constraint coming from $b\to s\gamma$ (at more than $95\%$
CL.). Ignoring this fact, we can accommodate the stops
near 600 GeV with the gluino weighing in at around 3 times that value,
i.e.~near 2 TeV.
Taking into account the $b\to s\gamma$ constraint at 2 sigma, pushes
up the stops to around 1000 GeV and correspondingly the gluino to
roughly 3 TeV.
Of course we should take this constraint seriously, however, the
reason for cutting it some slack is the general concern that
$\mu/B\mu$ in some cases might require some extra dynamics above or at
the messenger scale $M$, which could provide further contributions to
the Higgs masses. Indeed by pushing up the down Higgs $H_d$ it is
possible to crank up the heavy Higgses and hence the $H_\pm$ mass
beyond the flavor limit. Having this option in mind, we leave open
the possibility that the stops can be as light as 550 GeV.
The heavy squarks, viz.~those of the first and second generations are
commonly $5\div 17$ times heavier than the stops and hence rather safe
for flavor constraints such as those coming from $K-\bar{K}$,
$B-\bar{B}$, and $D-\bar{D}$ mixings.
Insisting that the model be free of further dynamics in the Higgs
sector, we can avoid the $b\to s\gamma$ limits by pushing up the
scales. In this case, interestingly, the $B\mu$ term can be generated
at the right order of magnitude as to have a reasonable
$\tan\beta\sim 10\div 25$. Since the $\mu$ term is generated
by the higher-dimension operator \eqref{eq:muterm}, this provides a
solution to the $\mu/B\mu$ problem at the price of the stops being
near or slightly above the TeV.

The predictions of the models are the following.
In this natural SUSic setting we can have the stop near $600\div 1100$ GeV.
The $\mu$ parameter is generically 100 GeV or so and there are
typically very light neutralini, chargini, staus and stau neutrino, in
the ballpark of $100\div 250$ GeV. Even though it might be some
challenge for the LHC this would be a thrill for the ILC.
The NLSP in the model can be either the lightest stau or the (mostly
Higgsino) neutralino.
Furthermore, this model giving rise to the flavor texture of the SM as
well as an inverted hierarchy of sparticles comes with $B'$, $W'$ and
$g'$ vector particles. The $W'$s are typically the lightest with a mass
of roughly
\beq
m_{W'} \gtrsim 2v \, ,
\eeq
where $v$ is typically near 2 TeV, and the saturation being at equal
$\SU(2)$ couplings of the two nodes $G_A,G_B$.

This concludes our overview of the model.
The reader interested in the details of the model is invited to
read on, while the others may jump to sec.~\ref{sec:benchmarks}.
In sec.~\ref{sec:unification}, we describe an extension of the model
which allows standard unification (as opposed to accelerated
unification which may be applied to the minimal model as well).

\section{Details of the minimal model\label{sec:minimalmodel}}

The model shown in fig.~\ref{fig:quiver2nodes} has qualitatively all
the ingredients for providing a successful phenomenology including a
relatively heavy Higgs particle -- at 126 GeV -- as well as a light
stop. We have in mind a low scale mediation scenario with the Higgsing
of the link fields taking place near the electroweak scale.
The model provides heavy first and second generation squarks as they
are situated close to SUSY breaking, while the third generation
squarks are light as they have suppressed masses due to the link
fields as in gaugino mediation. The third generation fermions
are heavy as they are placed on the same node as the Higgs fields
while the first and second generation fermions acquire masses via
higher dimension operators and hence are much smaller. Finally, in the
low scale mediation case, non-decoupled D-terms increase the
tree-level Higgs mass, alleviating the need for heavy stops or large
$A$-terms.

\subsection{Parameter space}

As mentioned in the introduction we seek to search for the lightest
possible stops in the parameter space of the above described
model. In order to cover as large a part of the parameter space as
possible, we invoke doublet-triplet splitting both in the messenger
sector and in the link sector. The parameter space is thus
parametrized in terms of the variables described below.
We consider a minimal messenger sector
\beq
\mathcal{W}_{T} = \sum_{i=1}^{n_{\rm mess}}
\left[S_2 T_{i2} \tilde{T}_{i2} + S_3 T_{i3} \tilde{T}_{i3}\right] \, ,
\qquad
\langle S_2\rangle = \eta M + \theta^2 F \, , \qquad
\langle S_3\rangle = M + \theta^2 F \, ,
\eeq
where the messengers $T_{i2},\tilde{T}_{i2}$ transform as
${\bf 2}_{-\frac{1}{2}},\bar{\bf 2}_{\frac{1}{2}}$ under
$\SU(2)_L\times \U(1)_Y$, respectively,
while $T_{i3},\tilde{T}_{i3}$ transform as
${\bf 3}_{-\frac{1}{3}},\bar{\bf 3}_{\frac{1}{3}}$ under
$\SU(3)_c\times \U(1)_Y$, respectively.\footnote{For sparticle spectra
with $x=F/M^2$ close to one, a coefficient in front of $F$ in $S_2$
cannot be rescaled into $\eta$, while for $x\lesssim 0.7$ the soft
masses do not change significantly and hence having both $\eta$ and
such a coefficient would be redundant. Here we fix the coefficient in
front of $F$ to be unity for the latter reason and for simplicity. }
The link field sector with doublet-triplet splitting provides the
following mass-squared matrix for the gauge bosons
\beq
\mathcal{M}^2_{v_k} = 2 v_{k}^2
\begin{pmatrix}
g_{A_k}^2 & -g_{A_k} g_{B_k} \\
-g_{A_k} g_{B_k} & g_{B_k}^2
\end{pmatrix}  \, ,
\eeq
with eigenvalues $0$ and
\beq
m_{v_k}^2 = 2\left(g_{A_k}^2 + g_{B_k}^2\right) v_k^2 \, ,
\eeq
where $k=1,2,3$ stand for $\U(1)$, $\SU(2)$ and $\SU(3)$,
respectively.
The link fields are bifundamental fields
$\{\omega_d,\omega_L\}$ as are their conjugates
$\{\tilde{\omega}_d,\tilde{\omega}_L\}$.
Here we are using the notation of \cite{Auzzi:2011eu} for the link
fields where $\omega_R,\tilde{\omega}_R$ denote a pair of fields
transforming under the representation $R,\bar{R}$ of $G_A$ and as
$\bar{R},R$ of $G_B$, respectively; $R$ is written in terms of a SM
field in such a representation. 
The VEVs of the link fields are
\beq
\langle\omega_d\rangle = \langle\tilde{\omega}_d\rangle = v_3 \, , \qquad
\langle\omega_L\rangle = \langle\tilde{\omega}_L\rangle = v_2 \, , \qquad
v_1^2 = \frac{3}{5}v_2^2 + \frac{2}{5}v_3^2 \, ,
\eeq
from which we define the parameters
\beq
y_k = \frac{m_{v_k}}{\sqrt{\eta}M} \, , \qquad
y \equiv \left(y_1y_2y_3\right)^{\frac{1}{3}} \, , \qquad
\kappa = \frac{v_2}{v_3} \, .
\eeq
At the Higgsing scale $m_{v_k}$ the standard model gauge couplings
$g_k$ are given by
\beq
\frac{1}{g_k^2} = \frac{1}{g_{A_k}^2} + \frac{1}{g_{B_k}^2} \, ,
\eeq
from which it is practical to define the following three angles
\beq
\tan\theta_k \equiv \frac{g_{A_k}}{g_{B_k}} \, .
\eeq
Finally, we define
\beq
x = x_3\equiv \frac{F}{M^2} \, , \qquad
x_2\equiv \frac{F}{\eta M^2} \, .
\eeq
In summary, the parameter space is parametrized by the set of variables
$\{M,x,y,\eta,\kappa,\theta_k,n_{\rm mess}\}$.

\subsection{SUSY-breaking masses\label{sec:SUSYmasses}}

The gaugino masses are similar to those of minimal gauge mediation
\cite{Martin:1996zb}, however with doublet-triplet splitting taken
into account,
\beq
m_{\tilde{g}_k} = n_{\rm mess} \left(\frac{g_k}{4\pi}\right)^2 \Lambda_k \,
q(x) \, ,  \qquad
q(x)=\frac{1}{x^2}\left[(1+x ) \log(1+x) + (1-x) \log(1-x) \right] \, ,
\label{mgauginos}
\eeq
where $n_{\rm mess}$ is the number of copies of messengers and the
effective SUSY breaking scales are given by
\beq
\Lambda_3 = \frac{F}{M} \, , \qquad
\Lambda_2 = \frac{F}{\eta M} \, , \qquad
\Lambda_1 = \frac{3}{5}\Lambda_2 + \frac{2}{5}\Lambda_3 \, .
\eeq
The sfermion masses are given in eq.~(4.2) of \cite{Auzzi:2011wt},
\begin{align}
m_{\tilde{f}_l}^2 &= 2 n_{\rm mess}
\bigg[
\left(\frac{g_{3}}{4\pi}\right)^4
\Lambda_3^2 \, C_{2,3}^{\tilde{f}} \,\mathcal{E}^l(x_3,y_3,\lambda_3)
+\left(\frac{g_{2}}{4\pi}\right)^4
\Lambda_2^2 \, C_{2,2}^{\tilde{f}} \,\mathcal{E}^l(x_2,y_2,\lambda_2)
\non & \phantom{= 2 n_{\rm mess} \bigg[}
+\left(\frac{g_{1}}{4\pi}\right)^4 \bigg(
\frac{2}{5}\Lambda_3^2 \, C_{2,1}^{\tilde{f}} \,\mathcal{E}^l(x_3,y_1,\lambda_1)
+\frac{3}{5}\Lambda_2^2 \, C_{2,1}^{\tilde{f}} \,\mathcal{E}^l(x_2,y_1\lambda_1)
\bigg) \bigg]
\ , \label{sferm2}
\end{align}
where $C_{2,k}^{\tilde{f}}$, $k=1,2,3$, is the quadratic Casimir of the
representation under which the sfermion ${\tilde f}$ transforms, while
the index $l$ runs over generations.
The function $\mathcal{E}^l$ for the first and second generations is
given by \cite{Auzzi:2011wt}
\begin{align}
\mathcal{E}^{1,2}(x,y,\lambda_k) &= \frac{1}{x^2}\bigg[
\alpha_0(x)
-\left(1-\lambda_k^2\right)\alpha_1(x,y)
-(1-\lambda_k)^2 y^2 \alpha_2(x,y)
-\frac{2(1-\lambda_k)}{y^2}\beta_{-1}(x)
+ \beta_0(x) \non &
\phantom{=\frac{1}{x^2}\bigg[\ }
+\frac{2(1-\lambda_k)}{y^2}\beta_1(x,y)
+(1-\lambda_k)^2\beta_2(x,y) \bigg] \, , \qquad
\lambda_k\equiv \frac{1}{\sin^2\theta_k} \, ,
\end{align}
whereas for the third generation it reads
\beq
\mathcal{E}^3(x,y) = \frac{1}{x^2} \left[
\alpha_0(x) - \alpha_1(x,y) - y^2 \alpha_2(x,y)
-\frac{2}{y^2}\beta_{-1}(x) + \beta_0(x) + \frac{2}{y^2} \beta_1(x,y)
+\beta_2(x,y) \right] \, .
\eeq
The $\alpha$s and $\beta$s are defined in app.~A of
\cite{Auzzi:2011wt}.
The soft mass of the link fields is also given by eq.~\eqref{sferm2},
with $\mathcal{E}^{\rm link}=\mathcal{E}^1+ \mathcal{E}^3$ and an
appropriate quadratic Casimir (see \cite{Auzzi:2011wt} for details).
Note that we work in part of the parameter space where $y\sim 1/100$
or so and hence the above formulae can to a good approximation be
described as $\mathcal{E}^{1,2}$ being that of minimal gauge mediation
\cite{Martin:1996zb} and $\mathcal{E}^3\sim 0$. In our studies we use
the full formulae even though the spectra obtained are not really sensitive
to the mentioned approximation.

\subsection{RG evolution}

In order to calculate particle spectra we need to evaluate the RG
running from the messenger scale -- which we take to be the geometric
average of that of the two messenger fields: $\sqrt{\eta} M$ -- down
to the Higgsing scale of the link fields $m_v\equiv y\sqrt{\eta}M$.
The beta function coefficients of the gauge couplings read
\begin{align}
b_{A_1} = 5 \, , \quad
b_{A_2} = -1 \, , \quad
b_{A_3} = -4 \, , \quad
b_{B_1} = \frac{33}{5} \, , \quad
b_{B_2} = 0 \, , \quad
b_{B_3} = -2 \, ,
\end{align}
while the beta functions for the masses in the model are given in
app.~\ref{app:betafunctions}.
In the above we have assumed that the doublet-triplet splittings in
the messenger sector and the link sector are small enough that running
from the average messenger scale to the average Higgsing scale is a
sufficiently good approximation.

\subsection{Threshold effects}

At scale $m_{v}$ the sfermion masses of the node $G_A$ (viz.~the
third generation ones) receive a contribution from integrating out the
link fields and the heavy gaugini
\cite{DeSimone:2008gm},
\begin{equation}
\delta m_{\tilde{f}}^{2}  = \sum_k \left(\frac{g_k}{4\pi}\right)^2 C_{\tilde{f},k}
\left[2\sin^2\theta_k(1-3\sin^2\theta_k) M_{k,B}^2
+ m_{v_k}^2 \tan^2\theta_k
\log\left(1+\frac{2m_{\omega_k}^2}{m_{v_k}^2}\right)
\right] \, , \label{eq:threshold_sfermions}
\end{equation}
where $m_{\omega_3}$ is the soft mass of $\omega_d,\tilde{\omega}_d$
and $m_{\omega_2}$ is that of $\omega_L,\tilde{\omega}_L$ while
$m_{\omega_1}^2 = \frac{3}{5} m_{\omega_2}^2 + \frac{2}{5} m_{\omega_3}^2$.
The soft masses of the Higgs fields at the scale $m_v$ receive the
following contribution,
\beq
\delta m_{H_{u,d}}^{2} = \delta m_{\tilde{L}}^{2}
+ \left(\frac{\lambda_{t,b} g_3}{4\pi^2}\right)^2
\left[2\sin^4\theta_3 M_{\tilde{g},B}^2
-\frac{1}{2}m_{v_3}^2\tan^2\theta_3
\log\left(1+\frac{2m_{\omega_3}^2}{m_{v_3}^2}\right) \right] \, ,
\eeq
where $\lambda_{t,b}$ are the Yukawa couplings of the top and bottom,
respectively.

\subsection{Higgs sector\label{sec:Dterms}}

In order to naturally acquire a Higgs mass of 126 GeV, we exploit the
fact that in the part of parameter space of interest, the D-terms do
not decouple completely in the presence of SUSY breaking,
\beq
V_{D} = \frac{g_2^2(1+\Delta_2)}{8}
\left|H_u^\dag\sigma^a H_u + H_d^\dag\sigma^a H_d\right|^2
+\frac{3}{5}\frac{g_1^2(1+\Delta_1)}{8}
\left|H_u^\dag H_u - H_d^\dag H_d\right|^2 \, ,
\eeq
where $\sigma^a$ are the Pauli matrices and
\beq
\Delta_k = \tan^2\theta_k \frac{2m_{\omega_k}^2}{m_{v_k}^2+2m_{\omega_k}^2} \, ,
\eeq
yielding tree-level Higgs masses
\cite{Batra:2003nj,Craig:2012hc}
\begin{align}
m^2_{h_0,H_0} &= \frac{1}{2}\left(m^2_{A_0} + \tilde{m}^2 \mp
\sqrt{(m^2_{A_0}-\tilde{m}^2)^2
+ 4\tilde{m}^2 m^2_{A_0}\sin^2(2\beta)}\right) \, , \\
m_{H^{\pm}}^2 &= m_{A_0}^2 + m_W^2 (1+\Delta_2) \, , \qquad
m_{A_0}^2 = 2|\mu|^2 + m_{H_u}^2 + m_{H_d}^2 \, ,
\end{align}
where the $\mu$ term is corrected as
\beq
|\mu|^2 = -\frac{1}{2}\tilde{m}^2
  - \frac{m_{H_u}^2 \tan^2\beta - m_{H_d}^2}{\tan^2\beta - 1} \, .
\eeq
The mass parameter $\tilde{m}$ is given by
\beq
\tilde{m}^2 = \frac{\frac{3}{5}g_1^2(1+\Delta_1)
+ g_2^2(1+\Delta_2)}{2} \, v_h^2 \, , \qquad
v_h = 174 \, {\rm GeV}\, ,
\label{mmvvhh}
\eeq
in terms of which the tree-level bound on the Higgs mass reads
\cite{Batra:2003nj,Maloney:2004rc,Craig:2011yk}
\beq
m_{h_0}^2 < \tilde{m}^2 \, .
\eeq
We furthermore assume that $B\mu$ is zero at the messenger scale $M$
and is generated by RG running
\beq
B\mu \simeq -\mu\left(\frac{3g_2^2}{8\pi^2} M_2 \log\frac{m_{v_2}}{M_2}
+ \frac{3g_1^2}{40\pi^2} M_1\log\frac{m_{v_1}}{M_1}\right) \, ,
\eeq
where $M_{1,2}$ are gaugino masses. By generating $B\mu$ dynamically
it is no longer possible to choose $\tan\beta$, which hence is
determined by 
\beq
\sin 2\beta = \frac{2B\mu}{m_{A_0}^2} \, .
\eeq
We denote by
\beq
r_x = \frac{g_{hxx}}{g_{hxx}^{SM}} \, , \qquad x=t, b, \tau, V, G, \gamma \, ,
\eeq
the effective Higgs couplings normalized to the respective SM one and
\beq
\mu_x = \frac{\sigma \times {\rm BR}(x)}{\sigma \times {\rm BR}(x)_{\rm SM}} \, ,
\qquad x=t, b, \tau, V, G, \gamma \, ,
\eeq
is the signal strength in each experimental channel.
The tree-level couplings (see e.g.~\cite{Djouadi:2005gj}) are rescaled
as
\beq
r_b=r_\tau=-\frac{\sin \alpha}{\cos \beta} \, , \qquad
r_t=\frac{\cos \alpha}{\sin \beta} \, , \qquad
r_V=\sin (\beta -\alpha) \, ,
\eeq
where the parameter $\alpha$ is defined as the mixing angle between
$(H_{0d}, H_{0u})$ as in the MSSM \cite{Martin:1997ns} and is given by
\beq
\tan 2\alpha =
\frac{m^2_{A_0}+\tilde{m}^2}{m^2_{A_0}-\tilde{m}^2}
\tan 2\beta \, .
\eeq
The corrections to $g_{h \gamma \gamma}$ and $g_{h G G}$ come from
one-loop diagrams; in the region of parameters studied in this paper,
the deviations from the standard model are quite negligible (see
\cite{Blum:2012ii,Espinosa:2012in} for a recent discussion).

The only Higgs couplings which can have a sizable modification are
\beq
r_b = r_\tau \approx  1 + 2\frac{\tilde{m}^2}{m^2_{A_0}} \, ,
\eeq
where the approximation is valid for large $\tan\beta$ and to the
leading order in $\tilde{m}/m_{A_0}$.
When $m_{H \pm}$ saturates the bound of 380 GeV from $b\to s\gamma$,
this gives $r_b,r_\tau\approx 1.2$.
This could enhance the signal strengths,
$\mu_b\simeq\mu_\tau\simeq 1.12$, which in turn
would suppress $\mu_\gamma\simeq 0.78$.
This is in some tension with current experimental
data, in which $h\to\gamma\gamma$ is enhanced
\cite{ATLAS-CONF-2012-093,CMS-PAS-HIG-12-020,Low:2012rj,Benbrik:2012rm,Corbett:2012dm,Giardino:2012dp,Ellis:2012hz,Espinosa:2012im,Carmi:2012in}.

\subsection{Constraints}

There are various constraints that we have to take into account in
finding viable spectra, which we now describe in turn. The
constraints come in two types; direct search bounds and indirect
limits such as the oblique parameters, other electroweak precision
tests (EWPTs) and flavor constraints.

In the class of direct constraints, we consider the bounds on the
first and second generation squarks as function of the gluino mass
\cite{ATLAS-CONF-2012-033}. For instance for a 1.5 TeV gluino, the
first and second generation squarks should be heavier than 1.5 TeV and
heavier than 1.75 TeV for a 1 TeV gluino (see fig.~7 in
\cite{ATLAS-CONF-2012-033}). Our spectra do in general obey these
constraints, so this particular constraint is not really limiting our
search.
For the stau and the mostly Higgsino neutralino, the only bounds we
can apply are due to LEP, hence $m_{\tilde{\tau}_1} > 82$ GeV and
$m_{\tilde{\chi}_1^0} > 46$ GeV. The latter is never needed for the
spectra at hand.

Searches for $\gamma\gamma+$MET put a bound on the chargino
$m_{\tilde{\chi}_1^{\pm}} > 270$ GeV \cite{Meade:2009qv,Ruderman:2011vv} in
the case of a mostly bino NLSP (the lightest neutralino being mostly
bino and the lightest chargino thus mostly wino). This situation
typically happens when $\mu$ is not sufficiently light, whereas when
$\mu <200$ GeV, both the lightest chargino and the
lightest neutralino are Higgsini and hence the NLSP is typically a
Higgsino neutralino. In this case the bound on the chargino that
applies is the LEP bound reading $m_{\tilde{\chi}_1^{\pm}} > 94$ GeV
\cite{Beringer:1900zz}.

Among the oblique parameters, $T$ is the important one and it receives
contributions from a diagram of Higgses exchanging a $\U(1)$ boson and
a triplet scalar coming from the bifundamental $\SU(2)$ link field
after it is Higgsed
(${\bf 2}\otimes {\bf 2} = {\bf 1}\oplus {\bf 3}$). This amounts to
\beq
\Delta T = \frac{v_h^2}{\alpha} \left[
\frac{3}{20} \frac{\sin^4\theta_1}{v_1^2}
+ \frac{g_{A_2}^4 v_2^2 \cos^2(2\beta)}
{\left(2(g_{A_2}^2+g_{B_2}^2)v_2^2 + 2m_{\omega_2}^2\right)^2} \right]
\, , \qquad v_h = 174 \; {\rm GeV} \, ,
\eeq
which by neglecting the first term and assuming
$|\cos(2\beta)|\simeq 1$, we can rewrite as
\beq
\left(1+\cot^2\theta_2\right)\frac{v_2}{v_h} +
\frac{m_{\omega_2}^2}{g_{A_2}^2 v_2 v_h} >
\frac{1}{\sqrt{4\alpha|\Delta T|}}
\simeq \frac{5.9}{0.07 + 0.08n} \, ,
\eeq
where the equality assumes a face value of $T=0.07$
\cite{Beringer:1900zz} and $n$ is the number of standard deviations
one wishes to allow. We choose to work with model points within
roughly $1.5\sigma$.

Other electroweak tests are relevant as well; while not expressed in
terms of oblique parameters, they are typically parametrized using a
(higher-dimension) operator basis, where the limits are applied to
the respective coefficients via a chi-squared fit to electroweak
precision data. The Lagrangian density of the higher-dimension
operators takes the form $\delta\mathcal{L} = a_X \mathcal{O}_X$, with 
$X$ being the operator in question.
The operators relevant here are \cite{Han:2004az,Han:2005pr}
$\mathcal{O}_h=|h_0^\dag D_\mu h_0|^2$,
$\mathcal{O}^t_{\psi\psi'}=(\bar\psi\gamma^\mu\sigma^a\psi)(\bar\psi'\gamma_\mu\sigma^a\psi')$,
$\mathcal{O}^t_{h\psi}=i(h_0^\dag\sigma^a D^\mu
h_0)(\bar\psi\gamma_\mu\sigma^a\psi)+{\rm h.c.}$,
$\mathcal{O}^s_{\psi\psi'}=(\bar\psi\gamma^\mu\psi)(\bar\psi'\gamma_\mu\psi')$,
$\mathcal{O}^s_{h\psi}=i(h_0^\dag D^\mu
h_0)(\bar\psi\gamma_\mu\psi)+{\rm h.c.}$,
with coefficients
\begin{align}
a_h &= -\frac{\alpha}{v_h^2} \Delta T \, , &\qquad
a^t_{\psi\psi'} &= -\frac{1}{8v_2^2}\cos^4\theta_2 \, , \\
a^t_{\psi\Psi'} &= a^t_{h \psi} =
  \frac{1}{8v_2^2}\sin^2\theta_2\cos^2\theta_2 \, , &\qquad
a^t_{h\Psi} &= - \frac{1}{8v_2^2}\sin^4\theta_2\, , \\
a^s_{\psi\psi'} &= -\frac{3}{10v_1^2}Y_\psi Y_{\psi'}
  \cos^4\theta_1 \, , &\qquad
a^s_{\psi\Psi'} &= \frac{3}{10v_1^2}Y_\psi Y_{\Psi'}
  \sin^2\theta_1\cos^2\theta_1 \, , \\
a^s_{h\psi} &= \frac{3}{20v_1^2}Y_\psi \sin^2\theta_1
  \cos^2\theta_1 \, , &\qquad
a^s_{h\Psi} &= -\frac{3}{20v_1^2}Y_\Psi \sin^4\theta_1 \, ,
\end{align}
where $\psi$ is a 1st or 2nd generation SM fermion, while $\Psi$ is a
third generation one. We use a chi-squared fit with data of
\cite{Han:2005pr} to limit the operator coefficients to within the
$3\sigma$ level.

A CMS search for neutral Higgs bosons decaying to tau pairs has been
able to exclude $A_0$ up to 450 GeV for $\tan\beta=45$, and 290 GeV for
$\tan\beta=20$, while for $\tan\beta$ below $7\div 8$ no additional
limit (to that of LEP) has been obtained, see fig.~3 in
\cite{Chatrchyan:2012vp}.
For $\tan\beta=7$ we required $m_{A_0} > 125$ GeV
while for $\tan\beta=20$, $m_{A_0} > 290$ GeV.

Constraints from $b\to s\gamma$, by comparing experiment to NNLO QCD (at
second order in the strong coupling), set the bound $m_{H_{\pm}} >
380$ GeV at $95\%$ CL \cite{Hermann:2012fc}. This new constraint
is a lot more restrictive than the former one
\cite{Misiak:2006zs}. The choice of conforming with the brand new
limit pushes up the spectra to some degree.
We checked, using the expressions in \cite{Bertolini:1990if}, that the
contributions to the $b\to s\gamma$ branching ratio mediated by
superpartners are negligible in the region of parameters relevant for
the benchmark points (the bound on $m_{H \pm}$ changes only at the
percent level).

The heavy gauge bosons may also mediate FCNC; the most dangerous
constraints come from $g'$, due to a stronger gauge coupling.
These contributions are suppressed by the small non-diagonal elements
of the matrices which diagonalize the Yukawa couplings in
eq.~\eqref{eq:Yukawas}; constraints from meson mixings are usually
satisfied for $v_3\gtrsim 2$ TeV with $\theta_3\approx \pi/4$.

\subsection{Benchmark points\label{sec:benchmarks}}

\begin{figure}[!tp]
\begin{center}
\mbox{
\subfigure{\includegraphics[width=0.48\linewidth]{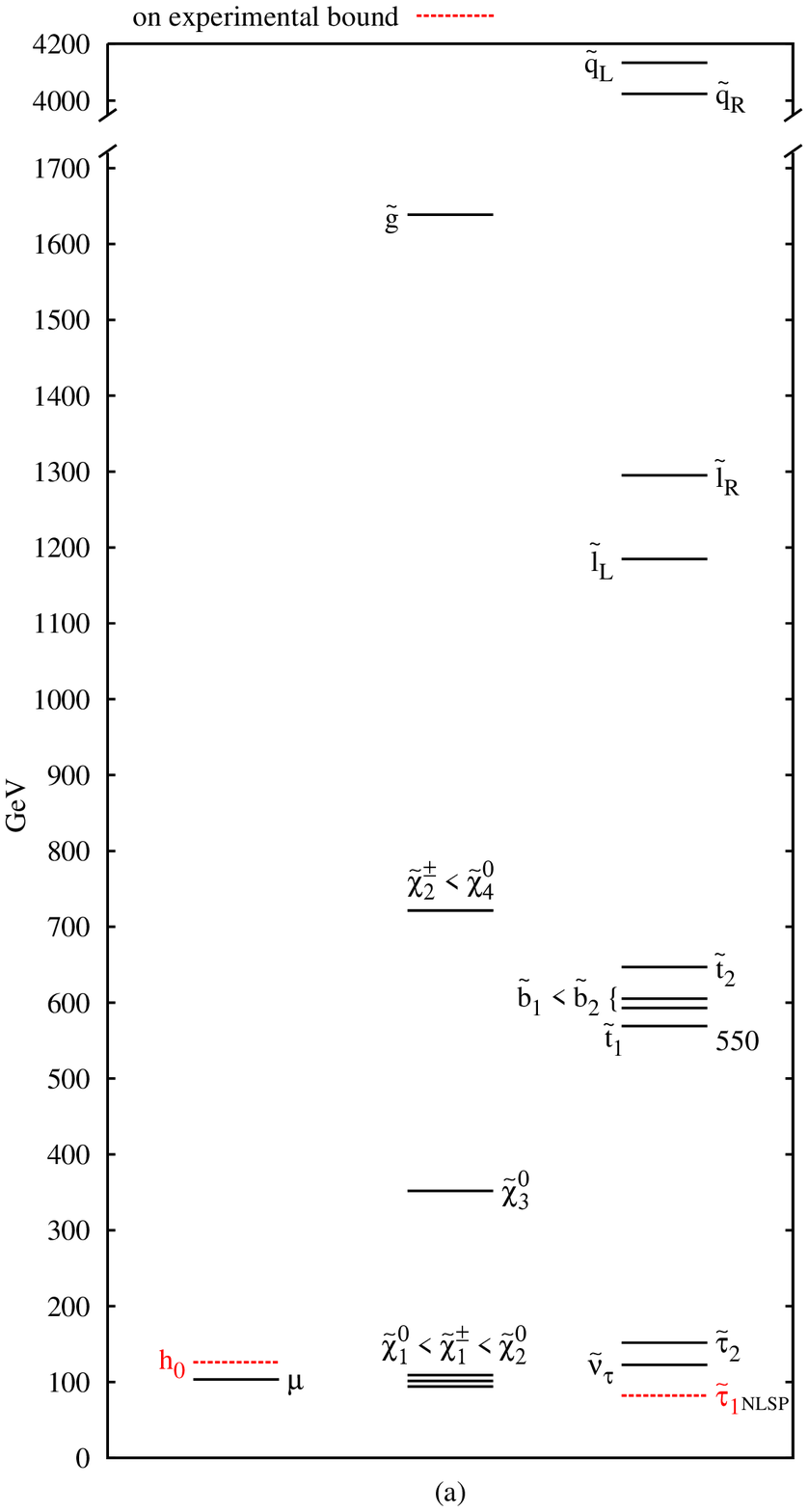}}\quad
\subfigure{\includegraphics[width=0.48\linewidth]{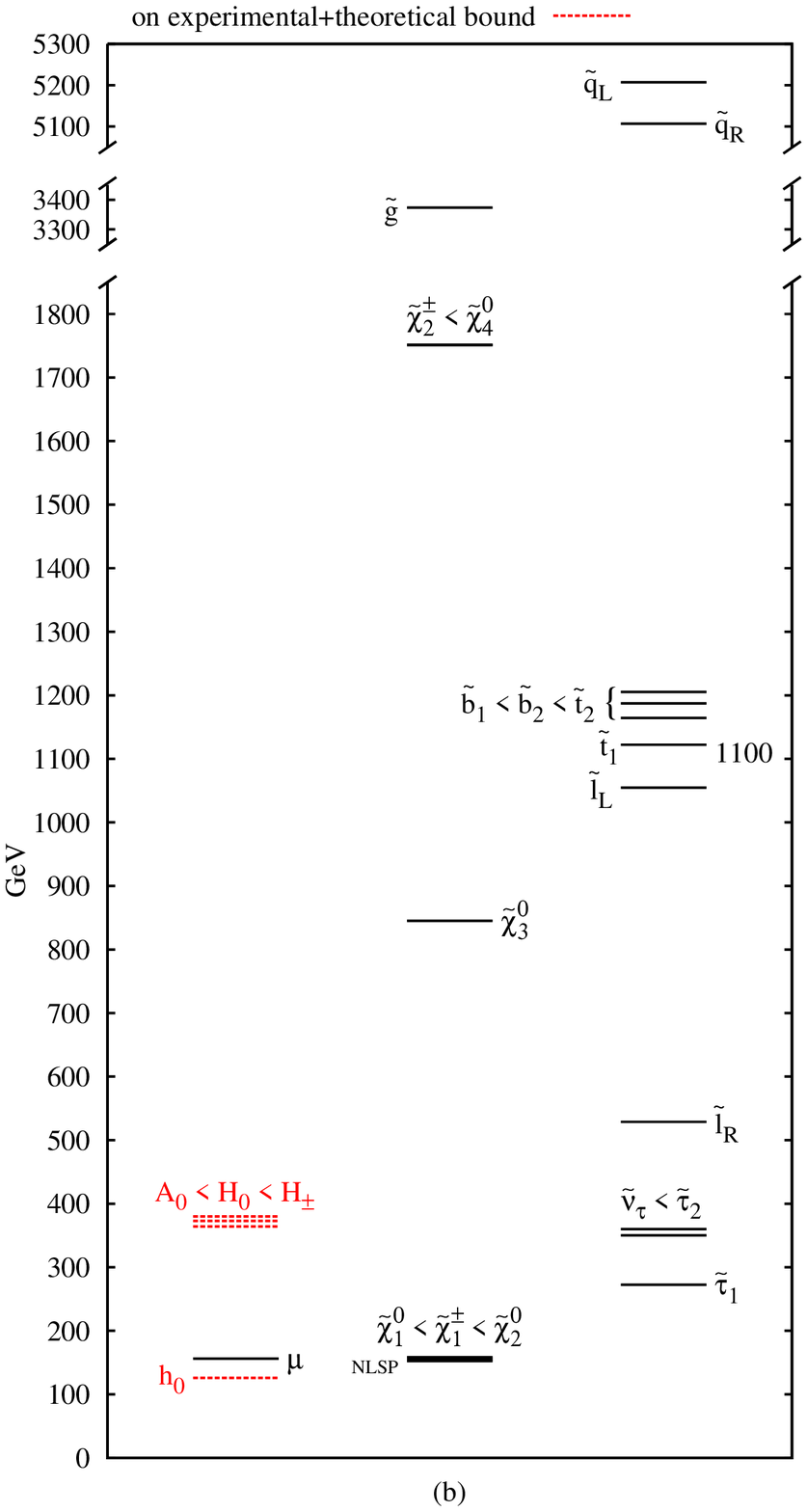}}}
\caption{(a) An example with a light stop; the heavy Higgs masses can
  be placed arbitrarily (in the range of $\sim 350\div 1000$ GeV) by
  choosing $m_{H_d}^2$ at the messenger scale $M$ -- this freedom has
  a negligible effect on the rest of the spectrum. 
  (b) An example of superpartners and Higgs masses in a self-contained
  minimal model -- including the Higgs sector all of whose soft
  masses, including $B\mu$, are dynamically generated -- that has a
  natural SM flavor texture; more details are given in
  appendix~\ref{app:benchmarkdetails}.
}
\label{fig:spectra10}
\end{center}
\end{figure}

Finally, we are ready to sum up the contributions to all the soft
masses described in secs.~\ref{sec:SUSYmasses}--\ref{sec:Dterms}, which
we plug into SOFTSUSY 3.3.0 \cite{Allanach:2001kg} in order to make
the final RG evolution from the scale $m_v$ down to the electroweak
scale, providing us with benchmark points describing the
characteristics of the model.
Fig.~\ref{fig:spectra10}a shows a benchmark point with a 126 GeV Higgs
and as light stop $\tilde{t}_1$ as we have been able to find in
parameter space.

Let us dwell a bit on the NLSP of the model under study.
Typically it is the (RH) stau or the Higgsino neutralino,
depending on the point in parameter space. Often the points with a
stau NLSP come hand in hand with a small VEV $v_2$ and
correspondingly a relatively light $W'$ which typically is at odds
with the EWPTs.
It is theoretically also possible that the stau neutrino is the NLSP,
which can happen also in a very small (and experimentally excluded)
corner of minimal gauge mediation. We find however that
all such points are excluded by EWPTs (maybe a particular corner is
still allowed by limits).

The mechanism giving rise to a light stau neutrino is the following.
In this model $L_3$ typically has a larger mass than $e_3$ and
this typically does not change even after the two-loop running from
the messenger scale down to the Higgsing scale. However, taking into
account the threshold effects of eq.~\eqref{eq:threshold_sfermions}
it is possible that the field $e_3$ receives a significantly larger
boost than $L_3$ when integrating out the link fields because
$m_{\omega_1}$ can be far larger than $m_{\omega_2}$. This is due to the
hypercharge squared of $e_3$ being four times bigger than that of
$L_3$ and hence if the threshold effect coming from integrating out
the link field $\omega_2$ is small enough, the RH stau can become
heavier than the LH stau and thus the stau neutrino can in principle
be the NLSP.

The heavier Higgses, $H_0,A_0,H_{\pm}$, are typically light in our
model as it stands.
Having light ``heavy Higgses'' (of the order of $\sim 350$ GeV) is a
prediction in the minimal incarnation of the model if we do not
allow additional dynamics in the Higgs sector to contribute to the
soft masses of the Higgses. This is constrained by $b\to s\gamma$ and
has consequences for the signal strength of Higgs decays.
In this minimal version of the model the stop typically needs to be
around the TeV in order for the spectrum to satisfy the constraints on
the Higgs sector (see fig.~\ref{fig:spectra10}b). 
If on the other hand we allow for additional contributions to the soft
masses of the Higgses at or above the messenger scale, then it is
possible to leave the stop as light as $\sim 550$ GeV (see
fig.~\ref{fig:spectra10}a). For
instance, increasing only the soft mass $m_{H_d}^2$ can push up the
``heavy Higgses'' above experimental bounds leaving the rest of the
spectrum more or less unchanged. 

Electroweak symmetry breaking (EWSB) is not per se an issue in the
model as it stands. However, the fact that all the soft masses of the
third generation as well as those of the Higgses start out negligible
at the messenger scale and acquire everything by RG evolution
constrains the gluino to weigh in at a certain level. This minimum
mass of the gluino, by means of the model, sets a lower bound on the
stops. We find that the lightest stop is typically heavier than $500$
GeV, consistent with the analysis of \cite{Craig:2012hc}. 
Notice that the gluino mass is also typically heavy in the model as it
is related to the soft masses of the first and second generation
squarks and link fields which need to be heavy to make $\Delta_{1,2}$
large and avoid the collider bounds. 
The most limiting constraint on the stop mass, however, comes from the
fact that it is correlated with the NLSP (often a RH stau) which has
to satisfy the LEP bound.

In order to allow for a natural texture in the fermion sector of the
model, we consider fixing the following parameters \cite{Craig:2011yk},
\beq
\epsilon_2 = \frac{v_2}{\Lambda_{\rm flavor}} = 0.02 \, , \qquad
\epsilon_3 = \frac{v_3}{\Lambda_{\rm flavor}} = 0.07 \, ,
\label{eq:texture}
\eeq
where $\Lambda_{\rm flavor}$ is the UV scale where flavor texture is
generated.
An inspection of the CKM matrix reveals that the $\epsilon$s
have to be large enough to reproduce the Cabibbo angle. If this is
not the case, the order one numbers of the higher-dimension operators
illustrated in eq.~\eqref{eq:Yukawas} have to be rather large.
As $y_k = m_{v_k}/M \simeq 3 v_k/M$ and it is required that
$\Lambda_{\rm flavor}\gtrsim M$, eq.~\eqref{eq:texture} puts a lower bound on
$y_k \gtrsim 3\epsilon_k$, with $k=2,3$. A spectrum with appropriately
chosen $y$s such as to allow for the above $\epsilon$s is shown in
fig.~\ref{fig:spectra10}b.

The example in fig.~\ref{fig:spectra10}b provides the spectrum of our
self-contained minimal model -- including the Higgs sector where all
its soft masses, including $B\mu$, are dynamically generated -- which
has a natural flavor texture and satisfies all direct as well as
indirect experimental bounds. 
The tuning of the Higgs mass-squared is at the percent level in this 
case.

\section{Extension with unification\label{sec:unification}}

\begin{figure}[!ht]
\begin{center}
\includegraphics[width=0.48\linewidth]{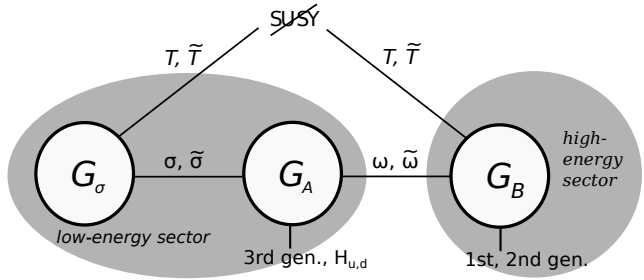}
\caption{A diagram describing the model with gauge groups
  $G_{\sigma}=\SU(2)$, $G_A=\U(1)\times \SU(2)\times \SU(3)$,
  $G_B=\SU(5)$
  and link fields $\sigma,\tilde{\sigma},\omega,\tilde{\omega}$.
  The $\SU(2)$ node is added to enhance the Higgs mass. }
\label{fig:quiver3nodes}
\end{center}
\end{figure}

We now present an extension of the minimal model described in
sec.~\ref{sec:minimalmodel}, which allows for gauge coupling
unification \cite{Batra:2003nj}.
The model is described by the quiver-like diagram in
fig.~\ref{fig:quiver3nodes}. 
The outline of the extended model is as follows.
The messenger scale is set near the GUT scale,
$M\sim M_{\rm GUT}$, and SUSY breaking is communicated via messengers
from a single secluded sector to both group $G_B=SU(5)$ and
$G_{\sigma}=SU(2)$ (hence the model has a single spurion).
The VEV of the link fields $\omega,\tilde{\omega}$ is
also taken to be near the GUT scale, such that $y_\omega$ is order 1; 
this is sufficient to generate the required inverted hierarchy in the
sparticle spectrum.
However, the soft masses of the $\omega$-type link fields are
negligible relative
to their Higgsing scale, and consequently their contribution to
$\Delta_{1,2}$ is negligible. Hence the VEV of the bifundamental link 
fields $\sigma,\tilde{\sigma}$ needs to be relatively near the
electroweak scale, namely $y_{\sigma}\sim 10^{-11}$ or 
smaller. This can in principle give rise to tachyonic
(LH) staus due to the large range of running of the
link fields $\sigma,\tilde{\sigma}$. The link fields need to be
sufficiently heavy in 
order for the $\Delta$s to be of order one, such that the lightest
CP-even Higgs mass can be placed near 125 GeV.
A counteracting mechanism is also at work, since by cranking up
$y_{\omega}$, the soft masses on the $G_A=G_{\rm SM}$ node are increased.
This in turn pushes up the stop mass and can then become a problem for
naturalness in the model. All this said, the model in principle
provides a viable unifying theory with a light stop and a 126 GeV
Higgs.

Let us comment on the possibilities for unification and flavor texture.
The link fields $\omega,\tilde{\omega}$ could be chosen to transform in the
${\bf 5},\bar{\bf 5}$ of $\SU(5)$ or alternatively in the
${\bf 10},\overline{\bf 10}$ which is much better for flavor physics
\cite{Auzzi:2011eu}. These representations will not prohibit the
gauge coupling unification of the group $G_A$ as they are complete
representations of $\SU(5)$ and also these links will run only a
little bit. The group $G_B$ is already chosen as an $\SU(5)$ and nothing
needs to be done here. 
One can further speculate on the unification of the ``low-energy
sector'' of fig.~\ref{fig:quiver3nodes}. The exceptional group $\E_6$
contains $\SU(2)\times\SU(6)$ which in turn contains
$\SU(2)\times\U(1)\times\SU(2)\times\SU(3)$. This is all what is
needed for the low-energy sector. It is also possible to consider the
so-called trinification group $\SU(3)^3$ which is a subgroup of
$\E_6$. This however requires the low-energy sector to be embedded in
$\E_8$ as it contains $\SU(3)\times\E_6$.

Finally, we make an estimate to see whether the many decades of running can
make the (LH) staus tachyonic. Using the two-loop beta functions of
app.~\ref{app:betafunctions} and the threshold corrections
\eqref{eq:threshold_sfermions}, taking into account the wino, the
heavy 1st and 2nd generation squarks as well as the link fields
$\sigma,\tilde{\sigma}$, we obtain the running mass for the (LH) stau
at the scale $m_{v_\sigma}$ (assuming it starts out vanishing)
\begin{align}
\delta m_{L_3}^2 &= \frac{3\alpha_2}{8\pi\cos^2\theta_\sigma}
\bigg[\Upsilon \log y_\sigma + \sin^2\theta_\sigma
\bigg] m_\sigma^2 \, ,
\label{eq:stau_tachyon_estimate} \\
\Upsilon &\equiv
\frac{\alpha_2}{\pi\cos^2\theta_\sigma}
\left(
\frac{1}{3}\left(\frac{4\alpha_3}{\alpha_2}\right)^2
\frac{\sin^4\theta_\sigma}{\sin^4\theta_{\omega,3}}
+4\frac{\tan^4\theta_\sigma}{\sin^4\theta_{\omega,2}} + 2
\right)
-\frac{8n_{\rm mess}}{3}\sin^4\theta_\sigma \, , \nonumber
\end{align}
where $g_\sigma = g_2 / \sin\theta_\sigma$, $m_{\sigma}^2$ is the
soft mass of the link fields $\sigma,\tilde{\sigma}$ and
$\alpha_k=g_k^2/(4\pi)$. We have neglected all contributions
proportional to $\alpha_1$ and we have assumed that the effective
SUSY-breaking scale $F/M$ is the same on both node
$G_\sigma$ and $G_B$.

\begin{figure}[!th]
\begin{center}
\includegraphics[width=0.5\linewidth]{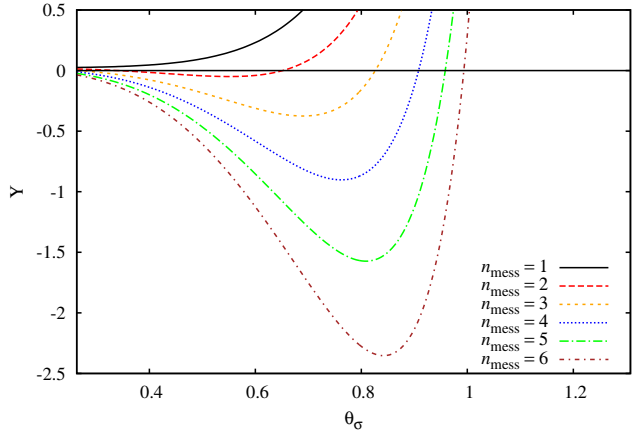}
\caption{$\Upsilon$ as function of $\theta_\sigma$ for various numbers
  of messengers $n_{\rm mess}$;
  here $\theta_{\omega,2}=\theta_{\omega,3}=\pi/4$. }
\label{fig:criticalthetas}
\end{center}
\end{figure}
Fig.~\ref{fig:criticalthetas} shows the value of $\Upsilon$ in
eq.~\eqref{eq:stau_tachyon_estimate} as function of
$\theta_\sigma$ and number of messengers $n_{\rm mess}$ for
$\theta_{\omega,2}=\theta_{\omega,3}=\pi/4$.
Whenever $\Upsilon$ is negative, any $y_{\sigma}<1$ (even parametrically
small) is free of problems with tachyons. The range for
$\theta_\sigma$ is chosen such that $\alpha_{\sigma},\alpha_{A+B,2}<1/2$
both remain perturbative.
For $n_{\rm mess}=1$, $\Upsilon$ is positive definite while for
$n_{\rm mess}>1$ it is negative for some range of $\theta_\sigma$. For
$\theta_\sigma=\pi/4$, $\Upsilon$ is negative for $n_{\rm mess}\geq 3$.

\section{Discussion}

In this paper we answered the question of how light the stop can be in
minimal supersymmetric quiver-like extensions of the SM,
which deal with {\it all} the eighteen SM parameters,
including a 126 GeV Higgs boson,
and which satisfy all current experimental bounds.
The answer depends on whether we allow for additional dynamics
modifying the soft masses of the Higgs sector or we assume
that the model be self contained.
If we allow modification of the Higgs masses we can accommodate the
stop near 550 GeV, while in a self-contained model as it stands,
the stop cannot be lighter than roughly a TeV.
In this latter version, the $B\mu$ term is radiatively generated due to heavy
electroweak gaugini allowing for a reasonably low $\tan\beta\sim 20$.
The heavy gaugini come along with a heavy gluino, and the latter
gives rise to some residual tuning.

We find that the properties of the spectra are rather robust.
A relatively light stop, near the $0.5\div 1$ TeV range,
is accompanied by a heavy gluino, with mass 
$m_{\tilde g}\sim 3m_{\tilde t}$, heavy 1st and 2nd generation
squarks, a factor of $3\div 20$ heavier than the stop, and a
relatively light $W'$, in the $3\div 10$ TeV range.
The NLSP is either a light Higgsino neutralino or a stau, near 100
GeV;
in the latter case, the $W'$ is lighter, and may be within the reach
of the LHC.
It may also be possible to obtain a stau neutrino NLSP in some corners
of parameter space,
though we did not manage to find an example that satisfies all our
constraints.

We have performed the search for the lightest possible third
generation squarks \emph{without} applying direct search limits to
them a priori. After we obtained the results we then checked whether
the stops or sbottoms (which are typically degenerate in our model)
are excluded or close to being discovered. 
In the regime of parameters studied in this paper,
the NLSP decays to a gravitino inside the detector; the direct search
limit (with only $2.05\; {\rm fb}^{-1}$ of integrated luminosity at
$\sqrt{s}=7$) for a 100 GeV neutralino and other colored sparticles
decoupled requires $m_{\tilde{t}_1}\gtrsim 270$ GeV
\cite{Aad:2012cz}.\footnote{In the case of a light bino even stronger
  limits exist \cite{Barnard:2012at}.} 
All our results comply with this limit, but the spectra with the
lightest stops (and additional contribution to the heavy Higgses)
could be discovered (or excluded) in the near future by the LHC.

The model as it stands is the minimal version and it does not allow
for a standard unification, though in some region of
parameter space it may allow for some type of accelerated unification
\cite{ArkaniHamed:2001vr,Craig:2012hc}. We have therefore contemplated
some extension with more gauge groups and link fields which may unify
in the standard way, to perhaps an $\E_n\times\E_m$ GUT.
We leave a detailed study thereof for the future.

Finally, let us discuss the predictions obtained via coupling to the
Higgs sector. 
In the case where the heavy Higgses are as light as allowed by direct
and indirect experimental constraints, i.e.~near 380 GeV, the
effective Higgs couplings to $\tau\bar\tau$ and $b\bar{b}$ are
enhanced by roughly $20\%$. Hence the signal strength in $h\to
b\bar{b},\tau\bar{\tau}$ increases by roughly $12\%$ which in turn
decreases that of $h\to\gamma\gamma$ by $22\%$.
An enhancement of the $h\to b\bar{b},\tau\bar{\tau}$ decay relative to
the $h\to\gamma\gamma$ 
one is thus a prediction of the model.
This may be in tension with the enhanced $h\to\gamma\gamma$ branching
ratio suggested by current experimental data
\cite{ATLAS-CONF-2012-093,CMS-PAS-HIG-12-020}.
However, the measurements are limited by significant theoretical
uncertainties in the calculation of the gluon fusion production cross
section and potentially also by experimental systematic errors
\cite{Baglio:2012et,Djouadi:2012rh}.
By increasing the masses of the heavy Higgses, the effective Higgs
couplings in our model become practically those of the SM.

\subsubsection*{Acknowledgments}

We thank Kfir Blum and Zohar Komargodski for discussions.
This work was supported in part
by the BSF -- American-Israel Bi-National Science Foundation,
and by a center of excellence supported by the Israel Science Foundation
(grant number 1665/10).
SBG is supported by the Golda Meir Foundation Fund.

\appendix

\section{Beta functions\label{app:betafunctions}}

The beta functions for the mass-squared of the sfermions
\cite{Martin:1993zk} in the model of sec.~\ref{sec:minimalmodel} are
given by
\beq
\frac{d m_{X}^2}{d\log\mu} = \frac{1}{(4\pi)^2} \beta_{X}^{(1)}
+ \frac{1}{(4\pi)^4} \beta_{X}^{(2)} \, ,
\eeq
where the coefficients for the particles of group $G_A$ are
\begin{align}
\beta_{Q_{3}}^{(1)} &= X_t + X_b \, , \quad
\beta_{Q_{3}}^{(2)} =
\frac{1}{6}S_{A}'
+ \left(\frac{1}{6}\right)^2 \sigma_{A_1}
+ \sigma_{A_2} + \sigma_{A_3} \, , \\
\beta_{u_{3}}^{(1)} &= 2X_t \, , \quad
\beta_{u_{3}}^{(2)} =
-\frac{2}{3}S_{A}'
+ \left(-\frac{2}{3}\right)^2 \sigma_{A_1}
+ \sigma_{A_3} \, , \\
\beta_{d_{3}}^{(1)} &= 2X_b \, , \quad
\beta_{d_{3}}^{(2)} =
\frac{1}{3}S_{A}'
+ \left(\frac{1}{3}\right)^2 \sigma_{A_1}
+ \sigma_{A_3} \, , \\
\beta_{L_{3}}^{(1)} &= X_{\tau} \, , \quad
\beta_{L_{3}}^{(2)} =
-\frac{1}{2}S_{A}'
+ \left(-\frac{1}{2}\right)^2 \sigma_{A_1}
+ \sigma_{A_2} \, , \\
\beta_{e_{3}}^{(1)} &= 2X_{\tau} \, , \quad
\beta_{e_{3}}^{(2)} =
S_{A}'
+ \sigma_{A_1} \, , \\
\beta_{H_u}^{(1)} &= 3X_t \, , \quad
\beta_{H_u}^{(2)} =
\frac{1}{2}S_{A}'
+ \left(\frac{1}{2}\right)^2 \sigma_{A_1}
+ \sigma_{A_2} \, , \\
\beta_{H_d}^{(1)} &= 3X_b + X_{\tau} \, , \quad
\beta_{H_d}^{(2)} =
-\frac{1}{2}S_{A}'
+ \left(-\frac{1}{2}\right)^2 \sigma_{A_1}
+ \sigma_{A_2} \, ,
\end{align}
while for the particles of group $G_B$ we have
\begin{align}
\beta_{Q_{1,2}}^{(1)} &= -\frac{32}{3}g_{B_3}^2|M_{B_3}|^2 -
6g_{B_2}^2|M_{B_2}|^2 - \frac{24}{5}\frac{1}{6}g_{B_1}^2|M_{B_1}|^2 \, , \quad
\beta_{Q_{1,2}}^{(2)} =
\frac{1}{6}S_{B}'
+ \left(\frac{1}{6}\right)^2 \sigma_{B_1}
+ \sigma_{B_2} + \sigma_{B_3} \, , \\
\beta_{u_{1,2}}^{(1)} &= -\frac{32}{3}g_{B_3}^2|M_{B_3}|^2
- \frac{24}{5}\left(-\frac{2}{3}\right)g_{B_1}^2|M_{B_1}|^2 \, , \quad
\beta_{u_{1,2}}^{(2)} =
-\frac{2}{3}S_{B}'
+ \left(-\frac{2}{3}\right)^2 \sigma_{B_1}
+ \sigma_{B_3} \, , \\
\beta_{d_{1,2}}^{(1)} &= -\frac{32}{3}g_{B_3}^2|M_{B_3}|^2
- \frac{24}{5}\frac{1}{3}g_{B_1}^2|M_{B_1}|^2 \, , \quad
\beta_{d_{1,2}}^{(2)} =
\frac{1}{3}S_{B}'
+ \left(\frac{1}{3}\right)^2 \sigma_{B_1}
+ \sigma_{B_3} \, , \\
\beta_{L_{1,2}}^{(1)} &= -6g_{B_2}^2|M_{B_2}|^2
- \frac{24}{5}\left(-\frac{1}{2}\right)g_{B_1}^2|M_{B_1}|^2 \, , \quad
\beta_{L_{1,2}}^{(2)} =
-\frac{1}{2}S_{B}'
+ \left(-\frac{1}{2}\right)^2 \sigma_{B_1}
+ \sigma_{B_2} \, , \\
\beta_{e_{1,2}}^{(1)} &=
- \frac{24}{5}g_{B_1}^2|M_{B_1}|^2 \, , \quad
\beta_{e_{1,2}}^{(2)} = S_{B}' + \sigma_{B_1} \, ,
\end{align}
and the link fields have
\begin{align}
\beta_{\omega_d}^{(1)} &=  -\frac{32}{3}g_{B_3}^2|M_{B_3}|^2
-\frac{24}{5}\frac{1}{3}g_{B_1}^2|M_{B_1}|^2\, , \non
\beta_{\omega_d}^{(2)} &=
-\frac{1}{3}S_{A}' + \frac{1}{3}S_{B}'
+ \left(-\frac{1}{3}\right)^2 \sigma_{A_1}
+ \left(\frac{1}{3}\right)^2 \sigma_{B_1}
+ \sigma_{A_3}
+ \sigma_{B_3} \, , \\
\beta_{\tilde{\omega}_d}^{(1)} &= -\frac{32}{3}g_{B_3}^2|M_{B_3}|^2
-\frac{24}{5}\left(-\frac{1}{3}\right)g_{B_1}^2|M_{B_1}|^2 \, , \non
\beta_{\tilde{\omega}_d}^{(2)} &=
\frac{1}{3}S_{A}' - \frac{1}{3}S_{B}'
+ \left(\frac{1}{3}\right)^2 \sigma_{A_1}
+ \left(-\frac{1}{3}\right)^2 \sigma_{B_1}
+ \sigma_{A_3}
+ \sigma_{B_3} \, , \\
\beta_{\omega_L}^{(1)} &= -6g_{B_2}^2|M_{B_2}|^2
-\frac{24}{5}\frac{1}{2}g_{B_1}^2|M_{B_1}|^2 \, , \non
\beta_{\omega_L}^{(2)} &=
-\frac{1}{2}S_{A}' + \frac{1}{2}S_{B}'
+ \left(-\frac{1}{2}\right)^2 \sigma_{A_1}
+ \left(\frac{1}{2}\right)^2 \sigma_{B_1}
+ \sigma_{A_2}
+ \sigma_{B_2} \, , \\
\beta_{\tilde{\omega}_L}^{(1)} &= -6g_{B_2}^2|M_{B_2}|^2
-\frac{24}{5}\left(-\frac{1}{2}\right)g_{B_1}^2|M_{B_1}|^2 \, , \non
\beta_{\tilde{\omega}_L}^{(2)} &=
\frac{1}{2}S_{A}' - \frac{1}{2}S_{B}'
+ \left(\frac{1}{2}\right)^2 \sigma_{A_1}
+ \left(-\frac{1}{2}\right)^2 \sigma_{B_1}
+ \sigma_{A_2}
+ \sigma_{B_2} \, .
\end{align}
We have defined the following symbols in the beta function
coefficients for group $G_A$
\begin{align}
X_t &\equiv 2\lambda_t^2 \left(m_{H_u}^2 + m_{\tilde{Q}_3}^2 +
m_{\tilde{u}_3}^2\right) \, , \\
X_b &\equiv 2\lambda_b^2 \left(m_{H_d}^2 + m_{\tilde{Q}_3}^2 +
m_{\tilde{d}_3}^2\right) \, , \\
X_{\tau} &\equiv 2\lambda_{\tau}^2 \left(m_{H_d}^2 + m_{\tilde{L}_3}^2 +
m_{\tilde{e}_3}^2\right) \, ,
\end{align}
where $\lambda_{t,b,\tau}$ are Yukawa couplings (we have neglected the
$A$-terms as they are not significant in our model) and
\begin{align}
S_{A}' \equiv \;&\frac{72}{25}g_{A_1}^4
\left[
\frac{1}{36} m_{\tilde{Q}_3}^2
-\frac{8}{9} m_{\tilde{u}_3}^2
+\frac{1}{9} m_{\tilde{d}_3}^2
-\frac{1}{4} m_{\tilde{L}_3}^2
+ m_{\tilde{e}_3}^2
+\frac{1}{4} m_{H_u}^2
-\frac{1}{4} m_{H_d}^2
\right] \non &
+\frac{18}{5} g_{A_1}^2 g_{A_2}^2
\left[
m_{\tilde{Q}_3}^2
- m_{\tilde{L}_3}^2
+ m_{H_u}^2
- m_{H_d}^2
\right]
+\frac{32}{5} g_{A_1}^2 g_{A_3}^2
\left[
m_{\tilde{Q}_3}^2
- 2 m_{\tilde{u}_3}^2
+ m_{\tilde{d}_3}^2
\right] \, , \\
\sigma_{A_1} \equiv \;&\frac{12}{25} g_{A_1}^4
\bigg[
m_{\tilde{Q}_3}^2
+ 8 m_{\tilde{u}_3}^2
+ 2 m_{\tilde{d}_3}^2
+ 3 m_{\tilde{L}_3}^2
+ 6 m_{\tilde{e}_3}^2
+ 3 m_{H_u}^2
+ 3 m_{H_d}^2 \non &
+ \sum_{R} \left(S_1(\omega_R) m_{\omega_R}^2
+ S_1(\tilde{\omega}_R) m_{\tilde{\omega}_R}^2\right)
\bigg] \, , \\
\sigma_{A_2} \equiv \;& 3 g_{A_2}^4
\left[
3 m_{\tilde{Q}_3}^2
+ m_{\tilde{L}_3}^2
+ m_{H_u}^2
+ m_{H_d}^2
+ \sum_{R} \left(S_2(\omega_R) m_{\omega_R}^2
+ S_2(\tilde{\omega}_R) m_{\tilde{\omega}_R}^2\right)
\right] \, , \\
\sigma_{A_3} \equiv \;&\frac{16}{3} g_{A_3}^4
\left[
2 m_{\tilde{Q}_3}^2
+ m_{\tilde{u}_3}^2
+ m_{\tilde{d}_3}^2
+ \sum_{R} \left(S_3(\omega_R) m_{\omega_R}^2
+ S_3(\tilde{\omega}_R) m_{\tilde{\omega}_R}^2\right)
\right] \, ,
\end{align}
while for the group $G_B$ we have
\begin{align}
S_{B}' \equiv \;&\frac{72}{25}g_{B_1}^4
\bigg[
\frac{1}{36} \left(m_{\tilde{Q}_1}^2 + m_{\tilde{Q}_2}^2\right)
-\frac{8}{9} \left(m_{\tilde{u}_1}^2 + m_{\tilde{u}_2}^2\right)
+\frac{1}{9} \left(m_{\tilde{d}_1}^2 + m_{\tilde{d}_2}^2\right)
-\frac{1}{4} \left(m_{\tilde{L}_1}^2 + m_{\tilde{L}_2}^2\right)
+ m_{\tilde{e}_1}^2 + m_{\tilde{e}_2}^2
\bigg] \non &
+\frac{18}{5} g_{B_1}^2 g_{B_2}^2
\left[
m_{\tilde{Q}_1}^2 + m_{\tilde{Q}_2}^2
- m_{\tilde{L}_1}^2 - m_{\tilde{L}_2}^2
\right] \non &
+\frac{32}{5} g_{B_1}^2 g_{B_3}^2
\left[
m_{\tilde{Q}_1}^2 + m_{\tilde{Q}_2}^2
- 2 \left(m_{\tilde{u}_1}^2 + m_{\tilde{u}_2}^2\right)
+ m_{\tilde{d}_1}^2 + m_{\tilde{d}_2}^2
\right] \, , \\
\sigma_{B_1} \equiv \;&\frac{12}{25} g_{B_1}^4
\bigg[
m_{\tilde{Q}_1}^2 + m_{\tilde{Q}_2}^2
+ 8 \left(m_{\tilde{u}_1}^2 + m_{\tilde{u}_2}^2\right)
+ 2 \left(m_{\tilde{d}_1}^2 + m_{\tilde{d}_2}^2\right)
+ 3 \left(m_{\tilde{L}_1}^2 + m_{\tilde{L}_2}^2\right)
+ 6 \left(m_{\tilde{e}_1}^2 + m_{\tilde{e}_2}^2\right) \non &
+ \sum_{R} \left(S_1(\omega_R) m_{\omega_R}^2
+ S_1(\tilde{\omega}_R) m_{\tilde{\omega}_R}^2\right)
\bigg] \, , \\
\sigma_{B_2} \equiv \;& 3 g_{B_2}^4
\left[
3 \left(m_{\tilde{Q}_1}^2 + m_{\tilde{Q}_2}^2\right)
+ \left(m_{\tilde{L}_1}^2 + m_{\tilde{L}_2}^2\right)
+ \sum_{R} \left(S_2(\omega_R) m_{\omega_R}^2
+ S_2(\tilde{\omega}_R) m_{\tilde{\omega}_R}^2\right)
\right] \, , \\
\sigma_{B_3} \equiv \;&\frac{16}{3} g_{B_3}^4
\left[
2 \left(m_{\tilde{Q}_1}^2 + m_{\tilde{Q}_2}^2\right)
+ \left(m_{\tilde{u}_1}^2 + m_{\tilde{u}_2}^2\right)
+ \left(m_{\tilde{d}_1}^2 + m_{\tilde{d}_2}^2\right)
+ \sum_{R} \left(S_3(\omega_R) m_{\omega_R}^2
+ S_3(\tilde{\omega}_R) m_{\tilde{\omega}_R}^2\right)
\right] \, .
\end{align}
We have neglected all Yukawa contributions at two loops for the
following reason. We anticipate an inverted hierarchy of sfermion
masses, hence we have neglected the first and second generation due to
small Yukawas (as usual) and the third generation is neglected not
because of the Yukawa but because the masses are assumed to be small
compared to the other contributions at two loop.
We have also neglected the contribution from the link fields to $S'_{A,B}$
as it is proportional to the difference in mass squared $m_{\omega}^2 -
m_{\tilde{\omega}}^2$ which in our model turns out to be very small (the
splitting is induced at two loops and it reaches a maximum of order 1
GeV at the end point of the running).
For the choice of link fields $\{\omega_d,\omega_L\}$ which corresponds to
the block diagonal parts of a ${\bf 5},\bar{\bf 5}$ bifundamental
field, the Dynkin indices read
\begin{align}
S_1(\omega_d) = S_1(\tilde{\omega}_d) = 6 \, , \qquad
S_2(\omega_d) &= S_2(\tilde{\omega}_d) = 0 \, , \qquad
S_3(\omega_d) = S_3(\tilde{\omega}_d) = 3 \, , \\
S_1(\omega_L) = S_1(\tilde{\omega}_L) = 6 \, , \qquad
S_2(\omega_L) &= S_2(\tilde{\omega}_L) = 2 \, , \qquad
S_3(\omega_L) = S_3(\tilde{\omega}_L) = 0 \, . \nonumber
\end{align}

\section{Details of benchmark points in fig.~\ref{fig:spectra10}\label{app:benchmarkdetails}}

\begin{center}
\begin{tabular}{|c|c|c|}
\hline
& Fig.~\ref{fig:spectra10}a & Fig.~\ref{fig:spectra10}b \\
\hline
input parameters &&\\
$M$ & $1.88\times 10^6$ GeV & $1.63\times 10^5$ GeV \\
$x$ & $0.115$ & $0.688$ \\
$y$ & $1/100$ & $1/11$ \\
$\eta$ & $0.79$ & $0.84$ \\
$\kappa$ & $0.13$ & $0.29$ \\
$(\theta_1,\theta_2,\theta_3)$ & $(0.69,1.3,\pi/4)$ & $(1.1,1.3,\pi/4)$\\
$n_{\rm mess}$ & 1 & 4\\
\hline
third generation &&\\
$m_{\tilde{t}_1}$ & $569$ GeV & $1122$ GeV \\
$\sqrt{m_{\tilde{t}_1}m_{\tilde{t}_2}}$ & $607$ GeV & $1163$ GeV\\
$m_{\tilde{\tau}_1}$ & $82$ GeV & $272$ GeV\\
\hline
lightest electroweakini &&\\
$\tilde{\chi}_1^0$ & $94$ GeV & $152$ GeV \\
$\tilde{\chi}_1^{\pm}$ & $101$ GeV & $156$ GeV \\
\hline
gluino and squarks &&\\
$m_{\tilde{g}} / m_{\tilde{t}_1}$ & $2.9$ & $3.0$ \\
$m_{\tilde{q}_L} / m_{\tilde{t}_1}$ & $7.3$ & $4.6$ \\
\hline
heavy vector bosons &&\\
$m_{B'}$ & $14.4$ TeV & $11.3$ TeV \\
$m_{W'}$ & $7.2$ TeV & $8.7$ TeV \\
$m_{g'}$ & $45.5$ TeV & $25.7$ TeV\\
\hline
Higgses &&\\
$m_{h_0}$ & $126$ GeV & $126$ GeV\\
$m_{H_0}$ & $-$ & $373$ GeV \\
$m_{A_0}$ & $-$ & $366$ GeV \\
$m_{H_\pm}$ & $-$ & $380$ GeV \\
$\mu$ & $103$ GeV & $156$ GeV \\
$B\mu$ & $-$ & $(87\;{\rm GeV})^2$\\
$\tan\beta$ & $25$ & $17.5$ \\
\hline
VEVs &&\\
$v_2$ & $2.0$ TeV & $2.7$ TeV \\
$v_3$ & $16.3$ TeV & $9.3$ TeV \\
\hline
\end{tabular}
\end{center}

\end{document}